\documentclass[]{jfm}
\usepackage{amsbsy}
\usepackage{epsfig}
\usepackage[english]{babel}
\usepackage{amssymb,amsmath,color}
\usepackage[mathscr]{eucal}
\usepackage{natbib}
\usepackage{microtype}
\usepackage[normalem]{ulem}

\usepackage{wry_abbreviationsHC}

\usepackage[usenames,dvipsnames]{xcolor}
\colorlet{cesar}{red}
\colorlet{bill}{blue}
\colorlet{stefan}{teal}

\usepackage{hyperref}
\hypersetup{
 breaklinks=true,
 colorlinks=true,
 linkcolor=blue,
 citecolor=Purple,
 urlcolor=Purple,
 pdfstartview={FitH},
 pdfview={FitH 0},
 pdfauthor={Cesar B. Rocha, Navid C. Constantinou, Stefan G. Llewellyn Smith, and William R. Young},
 pdftitle={The Nusselt numbers of horizontal convection},
 }

\newcommand{\ellx}{\ell_x}
\newcommand{\elly}{\ell_y}

\newcommand{\chidiff}{\chi_{\mathrm{diff}}}

\newcommand{\bs}{\ensuremath{b_{\mathrm{s}}}}

\newcommand{\bstar}{\ensuremath{b_{\star}}}

\newcommand{\Nu}{\mbox{\textit{Nu}}}
\newcommand{\Ra}{\mbox{\textit{Ra}}}

\renewcommand{\Pr}{\mbox{\textit{Pr}}}

\newcommand{\Nus}{\mbox{\textit{Nu}}_{\mathrm s}}
\newcommand{\NuF}{\mbox{\textit{Nu}}_{\mathrm F}}

\begin{document}

\title[The Nusselt numbers of horizontal convection]{The Nusselt numbers of horizontal convection}

\author{Cesar B.~Rocha\aff{1}
 \corresp{\email{crocha@whoi.edu}},
 Navid C.~Constantinou\aff{2},\\
 Stefan G.~Llewellyn Smith\aff{3,4},
 \and William R.~Young\aff{4}}

\affiliation{
\aff{1}Department of Physical Oceanography, Woods Hole Oceanographic Institution, Woods Hole, MA 02543, USA
\aff{2}Research School of Earth Sciences and ARC Centre of Excellence for Climate Extremes, Australian National University, ACT 2601, Australia
\aff{3}Department of Mechanical and Aerospace Engineering, University of California San Diego, 9500 Gilman Drive, La Jolla, CA 92093-0411, USA
\aff{4}Scripps Institution of Oceanography, University of California San Diego, 9500 Gilman Drive, La Jolla, CA 92093-0213, USA
}
%author{Cesar B. Rocha\aff{1}
% \corresp{\email{crocha@ucsd.edu}},
% Gregory L. Wagner\aff{2}
% \and William R. Young\aff{1}}
%
%\affiliation{\aff{1}Scripps Institution of Oceanography, University of California,
% San Diego
%\aff{2}Department of Earth, Atmospheric and Planetary Sciences, Massachusetts
% Institute of Technology}
%
%

\shortauthor{C.~B.~Rocha, N.~C.~Constantinou, S.~G.~Llewellyn Smith \& W.~R.~Young}

\pubyear{}
\volume{}
\pagerange{}
\date{ \today}
\setcounter{page}{1}

\maketitle

\begin{abstract}
We consider the problem of horizontal convection in which non-uniform buoyancy, $\bs(x,y)$, is imposed on the top surface of a container and all other surfaces are insulating. Horizontal convection produces a net horizontal flux of buoyancy, $\bJ$, defined by vertically and temporally averaging the interior horizontal flux of buoyancy. We show that $\overline {\bJ \bcdot \grad \bs}=- \kappa \la |\grad b|^2\ra$; the overbar denotes a space-time average over the top surface, angle brackets denote a volume-time average and $\kappa$ is the molecular diffusivity of buoyancy~$b$. This connection between $\bJ$ and $\kappa \la |\grad b|^2\ra$ justifies the definition of the horizontal-convective Nusselt number, $\Nu$, as the ratio of $\kappa \la |\grad b|^2\ra$ to the corresponding quantity produced by molecular diffusion alone. We discuss the advantages of this definition of $\Nu$ over other definitions of horizontal-convective Nusselt number currently in use. We investigate transient effects and show that $\kappa \la |\grad b|^2\ra$ equilibrates more rapidly than other global averages, such as the domain averaged kinetic energy and bottom buoyancy. We show that $\kappa \la |\grad b|^2\ra$ is essentially the volume-averaged rate of Boussinesq entropy production within the enclosure.  In statistical steady state, the interior entropy production is balanced by a  flux of entropy through the top surface. This leads to an equivalent ``surface Nusselt number'', defined as the surface average of vertical buoyancy flux  through the top surface  times the imposed surface  buoyancy $\bs(x,y)$. In experimental situations it is likely easier to evaluate the surface entropy flux, rather than the volume integral of $|\grad b|^2$ demanded by  $\kappa \la |\grad b|^2\ra$ . 

\end{abstract}

\section{Introduction}

Horizontal convection (HC) is convection generated in a fluid layer $0<z<h$ by imposing non-uniform buoyancy along the top surface $z=h$; all other bounding surfaces are insulated \citep{S08,R65,HG08}. HC is a basic problem in fluid mechanics and serves as an interesting counterpoint to the much more widely studied problem of Rayleigh--B\'enard convection (RBC) in which the fluid layer is heated at the bottom, $z=0$ and cooled the top, $z=h$.

In RBC the correct definition of the Nusselt number, $\Nu$, is clear: after averaging over~$(x,y,t)$ there is a constant vertical heat flux passing through every level of constant~$z$ between $0$ and $h$. By definition, the RBC $\Nu$ is the constant vertical heat flux through the layer divided by the diffusive heat flux of the unstable static solution.

In HC, however, there is zero net vertical heat flux through every level of constant~$z$. Thus, using notation introduced systematically in section~\ref{NussDef}, if the vertical flux of  buoyancy or heat  through the nonuniform surface is denoted by  $F(x)$ then $\overline{ F}=0$, where the overline denotes an $x$-average. To obtain a non-zero index of the  vertical  heat flux   \cite{R65,R98} defined the Nusselt number of horizontal convection  as a suitably normalized version of $\overline{|F|}$.  In section~\ref{NussDef} we discuss three other  definitions of the horizontal-convective $\Nu$, and recommend 
\beq
\Nu \defn {\chi}\slash{\chidiff}
\label{Sig1}
\eeq
as the best of the four. In~\eqref{Sig1}, $\chi$ is the dissipation of buoyancy (or thermal) variance defined in~\eqref{chidef} below and $\chidiff$ is the corresponding quantity of the diffusive (i.e.~zero Rayleigh number) solution. In the context of RBC, \cite{H63} remarked  that $\chi$ is a measure of the entropy production by thermal diffusion within the enclosure;  in this work we explore the  ramifications  of viewing~\eqref{Sig1} as a measure of HC entropy production.

The ratio on the right of~\eqref{Sig1} was introduced as a non-dimensional index of the strength of HC by \cite{PY02}. But because there seemed to be no clear connection to heat flux, \cite{PY02} did not refer to ${\chi}/{\chidiff}$ as a ``Nusselt number''. $\chi/\chidiff$ has been used as an index of HC in a few subsequent papers \citep{SKB04,RBLSY}. But most authors prefer to work with a  Nusselt number with a more obvious connection to the heat flux. In section~\ref{NussDef} we address this concern by establishing  relations between  $\Nu$ in~\eqref{Sig1} and the horizontal and vertical  heat fluxes in HC. This justifies referring to ${\chi}/{\chidiff}$ as a Nusselt number, and we note other advantages that  compel~\eqref{Sig1} as the best definition of a horizontal-convective Nusselt number.

Section~\ref{equilibration} discusses the transient adjustment of $\Nu$ in~\eqref{Sig1} to its long-time average and introduces a ``surface Nusselt number'', $\Nus$,  that is defined in terms of the entropy flux through the top surface. In statistical steady  state the surface flux of entropy  must  balance the  interior production of entropy and so, with sufficient time averaging, $\Nus=\Nu$. $\Nus$  has the advantage of requiring only a surface integral, rather than the volume integral of~$|\grad b|^2$ involved in~\eqref{Sig1}.  In section~\ref{equilibration} we also discuss the quantitative differences between the $\chi$-based $\Nu$ in~\eqref{Sig1} and Rossby's Nusselt number based on $\overline{|F|}$. Section~\ref{conclusion} is the conclusion.

\section{Formulation of the horizontal convection problem \label{form}}

 Consider a Boussinesq fluid with density $\rho=\rho_0(1 - g^{-1} b)$, where $\rho_0$ is a constant reference density, $b$ is the ``buoyancy,'' and $g$ is the gravitational acceleration. If the fluid is stratified by temperature variations then $b=g \alpha (T-T_0)$, where $T_0$ is a reference temperature and $\alpha$ is the thermal expansion coefficient. The Boussinesq equations of motion are
\begin{align}
\bu_t + \bu \bcdot \grad \bu + \bnabla p &= b \boldsymbol{\hat z} +
\nu \nabla^2 \bu \com \label{mom}\\
b_t + \bu \bcdot \grad b &= \kappa \nabla^2 b\com \label{buoy}\\
\bnabla \bcdot \bu &= 0 \per
\label{divu}
\end{align}
The kinematic viscosity is $\nu$ and the thermal diffusivity is $\kappa$.

\subsection{Horizontal convective boundary conditions and control parameters}

\begin{figure}
	\centering
	\includegraphics[width=\textwidth]{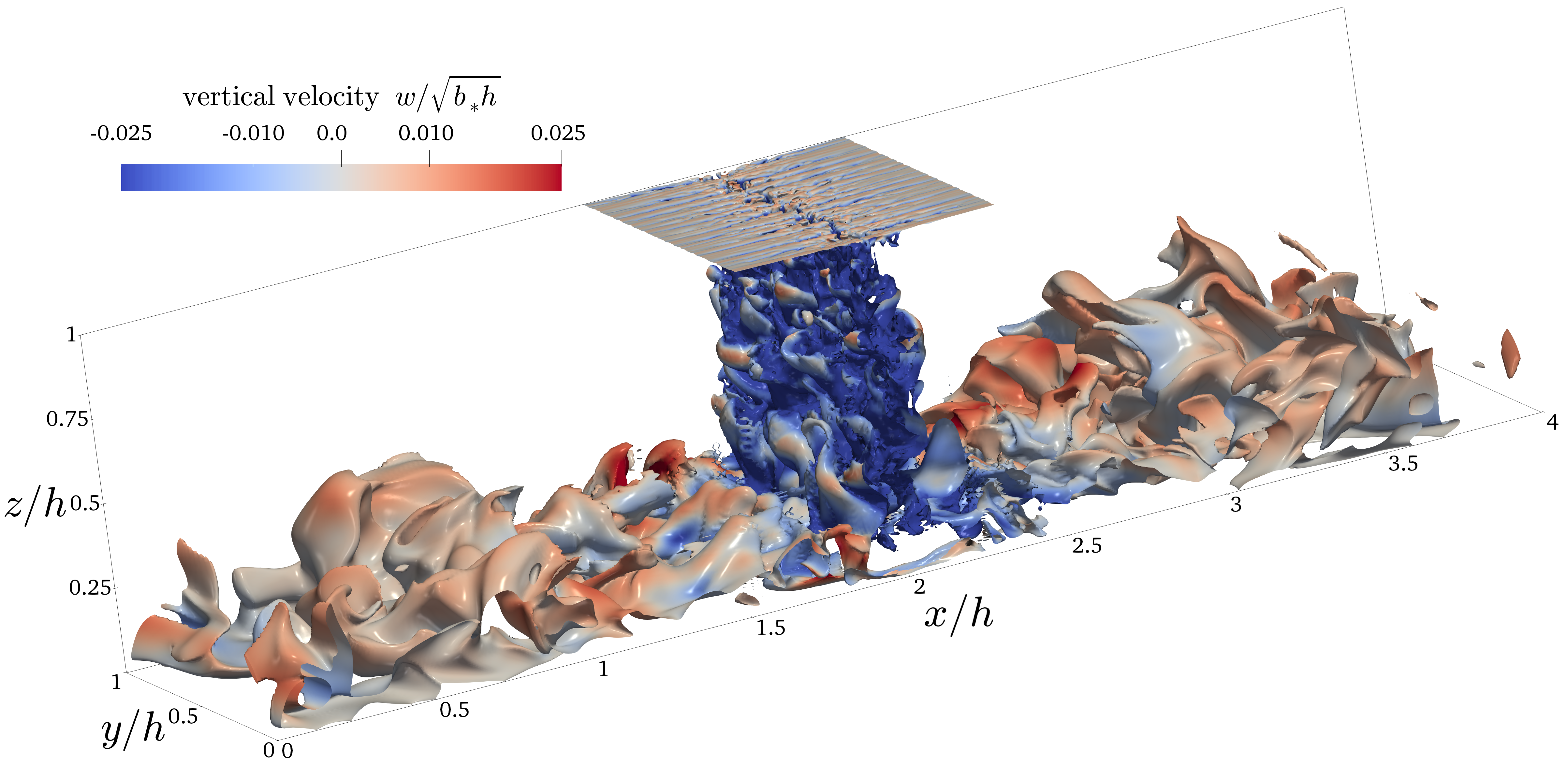}
	\caption{A snapshot of the $b=-0.7964\bstar$ surface at $\kappa t/h^2=0.12$ for a 3D horizontal-convective flow; colors denote the vertical velocity. This is a no-slip solution with sinusoidal surface buoyancy~$\bs$ in~\eqref{sbuoy8}; control parameters are $\Ra = 6.4 \times 10^{10}$, $\Pr=1$, $A_x= 4$ and $A_y=1$.}
	\label{Fig1}
\end{figure}

\begin{figure}[h]
	\centering
	\includegraphics[width=0.9\textwidth]{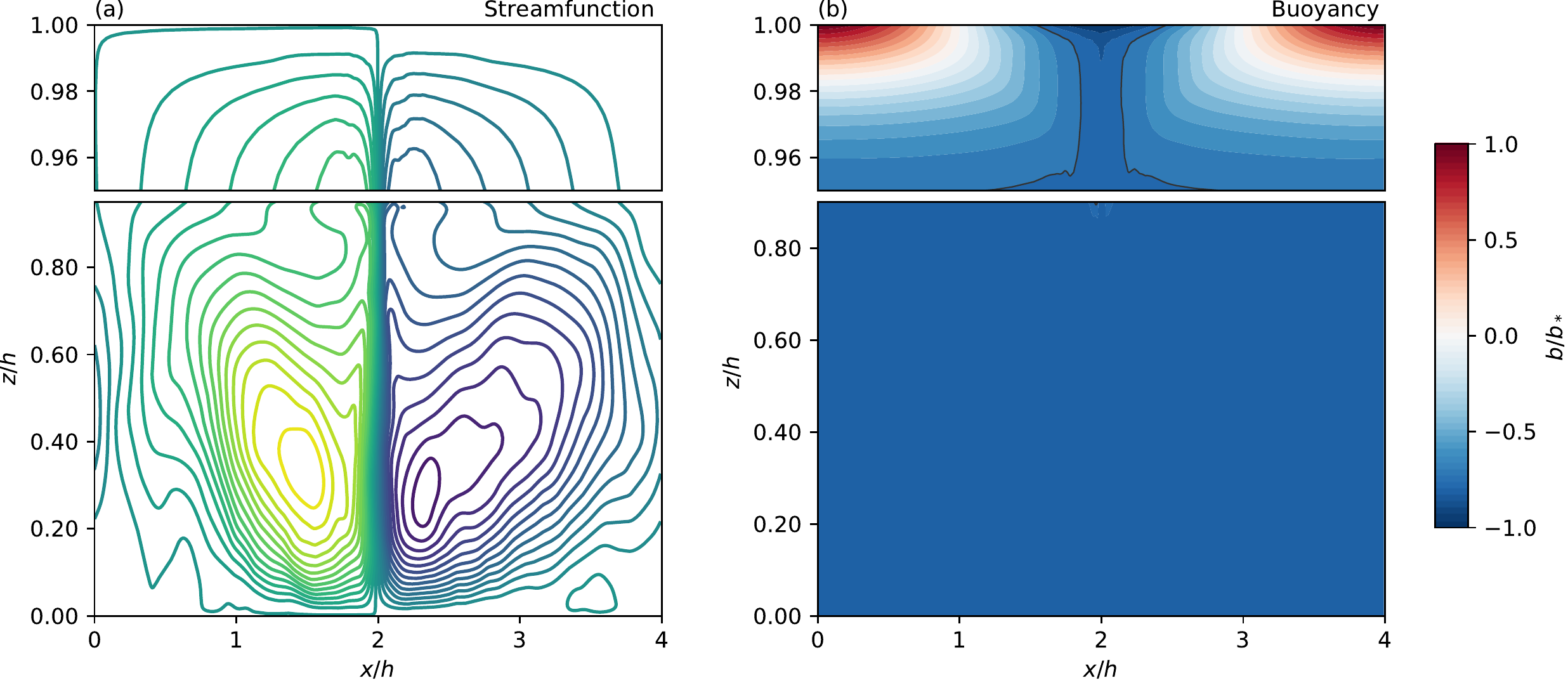}
	\caption{Snapshots of (a) streamfunction, calculated from the $y$-averaged velocity ($u$, $w$), and (b)~the $y$-averaged buoyancy at $\kappa t/h^2=0.12$. This is the same solution as that of figure~\ref{Fig1}. The streamfunction in panel~(a) is defined by the $y$-average of $u$ and $w$. The black contour in panel~(b) is $b =-0.79 \bstar$, which is close to the bottom buoyancy, defined as the $(x,y,t)$-average of $b$ at $z=0$.}
 	\label{Fig2}
\end{figure}

We suppose the fluid occupies a rectangular domain with depth $h$, length $\ellx$, width~$\elly$; we assume periodicity in the horizontal directions, $x$ and $y$. At the bottom surface ($z=0$) and top surface ($z=h)$ the boundary conditions on the velocity $\bu=(u,v,w)$ are $w=0$ and for the viscous boundary condition either no slip, $u=v=0$, or free slip, $u_z=v_z=0$. At $z=0$ the buoyancy boundary condition is no flux, $\kappa b_z=0$ and at the top, $z=h$, the boundary condition is $b = \bs(x)$, where the top surface buoyancy $\bs$ is a prescribed function of~$x$. As a surface buoyancy field we use
\beq
\bs(x) = \bstar \cos (k x) \com
\label{sbuoy8}
\eeq
where $k \defn 2 \pi/\ellx$.
Figure~\ref{Fig1} shows a 3D HC flow with no-slip boundary conditions and the
surface buoyancy~\eqref{sbuoy8}. Figure~\ref{Fig2} shows $y$-averaged buoyancy
and the overturning streamfunction calculated by the $y$-averaging the 3D
velocity. These figures illustrate three main large-scale features of HC: a
buoyancy boundary layer pressed against the non-uniform upper surface and
uniform buoyancy in the deep bulk; an entraining plume beneath the densest
point on the upper surface; and interior upwelling towards the nonuniform surface in the bulk of the domain.

% \cite{IV12} emphasize that the deep eddying flow takes place in a largely unstratified fluid and produces little interior mixing. This observation is important in our later discussion of horizontal-convective buoyancy flux.

The problem is characterized by four non-dimensional parameters: the Rayleigh and Prandtl numbers
\beq
\Ra \defn \frac{\ellx^3 \bstar }{\nu \kappa}\com \qquad \text{and} \qquad \Pr \defn \frac{\nu}{\kappa}\com
\label{controlDef1}
\eeq
and the aspect ratios $A_x \defn {\ellx}/{h}$ and $A_y \defn {\elly}/{h}$. Two-dimensional HC corresponds to $A_y=0$.

\subsection{Horizontal-convective power integrals}

We use an overline  $\overline{\,\cdots\,}$  to denote an average over $x$, $y$ and~$t$, taken at any fixed $z$ and angle brackets $\la \,\cdots \,\ra$ to denote a total volume average over $x$, $y$, $z$ and~$t$. Using this notation, we recall some results from \cite{PY02} that are used below.

Horizontally averaging the buoyancy equation~\eqref{buoy} we obtain the zero-flux constraint
\beq
\overline{wb} - \kappa \bar b_z=0\per
\label{PY1}
\eeq
Forming $\la \bu\bcdot\eqref{mom}\ra$, we obtain the kinetic energy power integral
\begin{align}
 \varepsilon = \la w b\ra \com \label{PY2}
\end{align}
where $\varepsilon \defn \nu \la |\grad \bu|^2\ra$ is the rate of dissipation of kinetic energy and $\la wb\ra$ is rate of conversion between potential and kinetic energy.

Vertically integrating~\eqref{PY1} from $z=0$ to $h$ we obtain another expression for $\la wb\ra$; substituting this into~\eqref{PY2} we find
\begin{align}
 \varepsilon = \kappa\, \Delta \bar b/h\com \label{PY2.5}
\end{align}
where $\Delta \bar b = \bar b(h) - \bar b(0)$ is the difference between the $(x,y,t)$-average of the buoyancy at the top, $z=h$, and the buoyancy at the bottom, $z=0$. The buoyancy difference can be bounded using the extremum principle for the buoyancy advection-diffusion equation~\eqref{buoy}. In the case of the sinusoidal profile in~\eqref{sbuoy8}, with $\bar b(h)=0$,  this leads to the inequality
\beq
\varepsilon \leq \kappa\, \bstar/h \per \label{PY3}
\eeq
In the example shown in figure~\ref{Fig1} the bottom buoyancy is $\bar b(0) \approx -0.83 \bstar$ and thus the right of inequality~\eqref{PY3} is about 20\% larger than~$\varepsilon$.

\section{Definition of the horizontal-convective Nusselt number \label{NussDef}}

For equilibrated HC the vertical buoyancy flux is zero through every level --- see~\eqref{PY1} --- and cannot be used to define a Nusselt number analogous to that of RBC. Moreover, with specified $\bs(x)$, buoyancy is transported along the $x$-axis and it is natural to consider the net horizontal flux,
\beq
J(x) \defn \frac{1}{\tau h \elly} \int_0^\tau \int_0^{\elly} \!\!\!\int_{0}^h (u b - \kappa b_x) \, \dd z \dd y \dd t \com
\label{Jdef}
\eeq
in defining an HC Nusselt number. In~\eqref{Jdef} $\tau$ is a time horizon sufficiently long so that the time average removes unsteady fluctuations remaining after the spatial average over the $(y,z)$-plane. But $J(x)$ is not constant and it is not initially obvious how to best define a single number out of $J(x)$ as an index of HC buoyancy transport.

\begin{figure}
 \centering
 \includegraphics[width=0.9\textwidth]{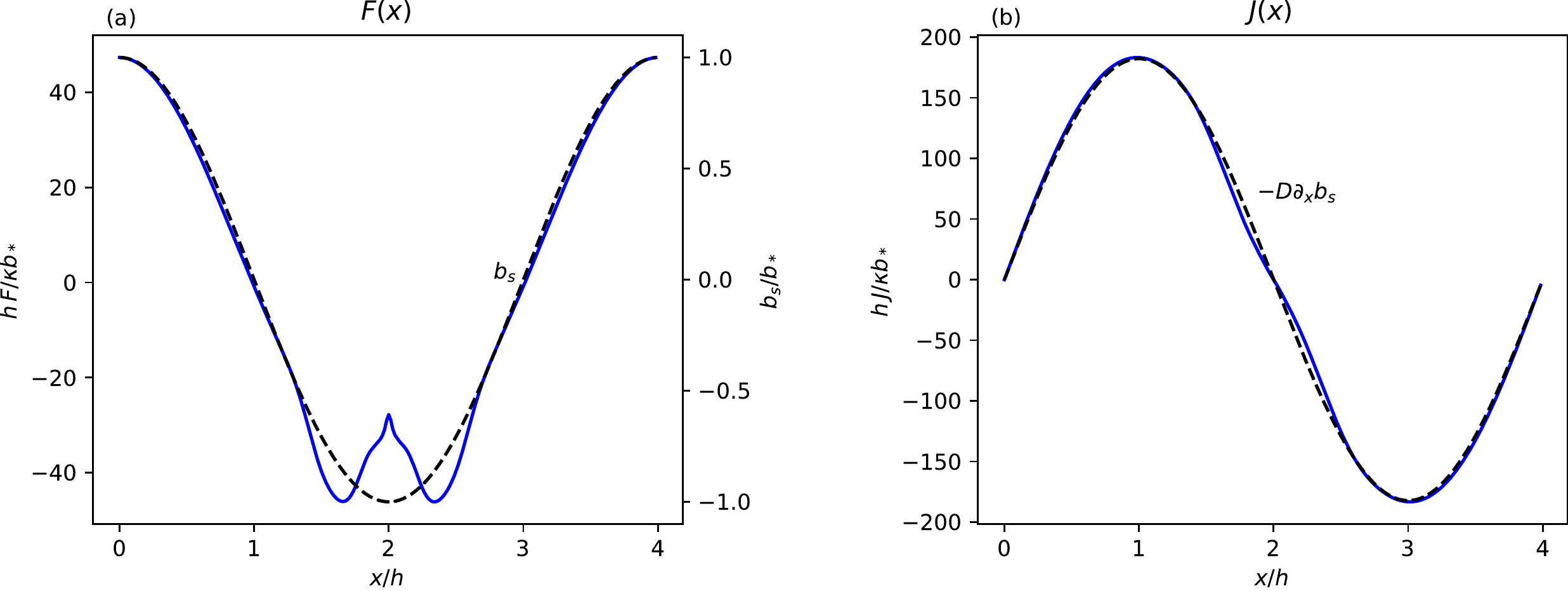}
 \caption{Left panel (a): the dashed black curve is the sinusoidal surface buoyancy profile in~\eqref{sbuoy8} and the solid blue curve is the vertical flux $F(x)$ defined in~\eqref{fdef}. Right panel (b) shows the horizontal flux $J(x)$ defined in~\eqref{Jdef} (the solid blue curve) and the relation~\eqref{diffFlux} with $D$ diagnosed from~\eqref{enhanceRat} (the dashed black curve). This is the same solution as that of figures~\ref{Fig1} and~\ref{Fig2}.}
 \label{Fig3}
 \end{figure}

Another measure of  horizontal-convective transport is provided by the averaged vertical buoyancy flux through the non-uniformly heated surface at $z=h$:
\beq
F(x) \defn \frac{1}{\tau \elly} \int_0^\tau \int_0^{\elly} \!\!\! \kappa b_z(x,y,h,t) \, \dd y \dd t \per
\label{fdef}
\eeq
Averaging the buoyancy equation~\eqref{buoy} over the $(y,z)$-plane, and also over time, one finds that the divergence of the $x$-flux in~\eqref{Jdef} is equal to the flux in and out through the top:
\beq
h \frac{\dd J}{\dd x}=F\per
\label{J1}
\eeq
Figure~\ref{Fig3} exhibits the functions $J(x)$ and $F(x)$ for the $\Ra = 6.4\times 10^{10}$ solution shown in figure~\ref{Fig1}.

\subsection{Two horizontal convective Nusselt numbers}

Following \cite{R65,R98}, some authors use,  more or less, the Nusselt number 
\beq
\NuF \defn \overline{|F|}\slash \overline{|F_{\text{diff}}|}\per
\label{NuF}
\eeq
In~\eqref{NuF}, $F_{\text{diff}}$ is the vertical flux of the diffusive solution i.e.~the vertical buoyancy flux produced by the solution of $\nabla^2 b_{\text{diff}}=0$, with $b_{\text{diff}}$ satisfying the same boundary conditions as $b$  \citep[e.g.][]{CWHL08,TsaiSheard2020}. We say ``more or less''  because in practice the normalizing denominator $\overline{|F_{\text{diff}}|}$ in~\eqref{NuF} is  sometimes replaced with a scale estimate such as $\kappa \bstar/\ellx$. 

As a variant of~\eqref{NuF}, other authors define a Nusselt number based on the net buoyancy flux through that part of the non-uniformly heated surface with a destabilizing vertical buoyancy flux i.e.~that part of the surface where $F(x)<0$ \citep[e.g.][]{HG08,G13,G14,Rose17}. Let  $X^-$ denote that part of the non-uniformly heated surface where $F=-|F|<0$, and $X^+$ the part where $F=+|F|>0$. Because there is no net flux through the surface \beq
\int_{X_-} \!\!\! F(x)\,  \dd x + \int_{X_+} \!\!\! F(x)\,  \dd x=0\com \label{mess}
\eeq
and so
\beq
\int_{X_-} \!\!\! |F(x)|\,  \dd x = \int_{X_+} \!\!\! F(x)\,  \dd x = \half \ellx \overline{|F|}\per
\label{mess3}
\eeq
Despite the close connection between the relation above and the numerator on the right of~\eqref{NuF}, this alternative Nusselt number is not simply related to $\NuF$: one must  normalize $X_-$  integral in~\eqref{mess3} by the length of the interval $X_-$: this unknown length  varies with $\Ra$ and  is not equal to $\ellx/2$ --- though in figure~\ref{Fig3}(a) it is close. Hence \textit{a priori} there is not a simple relation between $\NuF$ and the Nusselt number based on destabilized part of the boundary. 

% \vskip 0.1truein
% {\color{red}[NCC: I have a question: is the length of the $X_+$ interval then exactly $\ellx/2$? Because this is what 3.6 suggests... right?]} \textcolor{Mahogany}{WRY: No the length of $X_+$ does not have to be  exactly $\ellx/2$.:  the $X_+$ region, with $F>0$, might occupy 10\% of the domain, and still the integrals of $F$ over $X_+$ and $X_-$ are equal in absolute value,  but opposite  in sign --- that's all (3.6) says.}
% \vskip 0.1truein

\cite{SKB04} considered, and rejected, the  Nusselt numbers discussed above as too difficult for theoretical work. The problem is that one does not know in advance where buoyancy flows into and out of the domain and it is difficult to get an analytic grip on $|F|$. As an indication of the difficulty, there is no proof that the two $F$-based Nusselt numbers discussed above, when correctly normalized by the diffusive solution, are strictly greater than one. One expects on physical grounds that any fluid motion must increase buoyancy transport above that of the diffusive solution. A good definition of Nusselt number should manifestly satisfy this basic requirement and the Nusselt numbers above do not.

\subsection{The Nusselt number $\Nu \defn{\chi}/{\chidiff} $ and some of its properties}

The alternative to the Nusselt numbers discussed above is to use the diffusive dissipation of buoyancy variance,
\beq
\chi \defn \kappa \la |\bnabla b|^2\ra\com
\label{chidef}
\eeq
and define the Nusselt number as in~\eqref{Sig1}
\citep{PY02,SKB04,RBLSY}. It is straightforward to show that the $\Nu$ in~\eqref{Sig1} is greater than unity: amongst all functions $c(\bx)$ that satisfy the same boundary conditions as $b(\bx,t)$, the diffusive solution $b_{\text{diff}}(\bx) $ minimizes the functional $\la |\grad c|^2\ra$.

Now $\chi$ in~\eqref{chidef} does not have an obvious connection to the fluxes $J(x)$ and $F(x)$ in~\eqref{Jdef} and~\eqref{fdef}. This is probably why ${\chi}/{\chidiff}$ has not been popular as an index of the strength of HC. \cite{SKB04} refer to $\chi$ as a ``pseudo-flux'' because $\chi$ seems not to have a clear connection to the horizontal flux $J(x)$. But in~\eqref{justify} below we establish an integral relation between $\chi$ and $J$: this supports ${\chi}/{\chidiff}$ as a useful definition of $\Nu$.

Multiplying the buoyancy equation~\eqref{buoy}  with $b$ and taking the total volume-time average $\la \ra$ we obtain a ``power integral'' that expresses  $\chi$ entirely in terms of conditions at the nonuniform surface $z=h$:
\beq
\chi =  \overline{F\bs}\slash h  \com
 \label{chi7}
\eeq
where $F(x)$ is the vertical buoyancy flux in~\eqref{fdef} and $\bs(x)$ is the nonuniform surface buoyancy. Noting that $\chi>0$, we see that~\eqref{chi7} has an intuitive physical interpretation: on average buoyancy enters the domain ($F(x)>0$) where $\bs(x)$ is higher than its average value and leaves ($F(x)<0$) at locations where $\bs(x)$ is lower than its average value. In section~\ref{surfNuss} we discuss further useful properties of the buoyancy power integral~\eqref{chi7}.

Using~\eqref{J1} to replace $F$ in~\eqref{chi7} by $\dd J/\dd x$, and integrating by parts in $x$, one finds
\beq
\chi = - \overline{J\frac{\dd \bs}{\dd x}} \per
 \label{justify}
\eeq
Thus $\chi$ is directly related to an $x$-average of $J(x)$, weighted by the surface buoyancy gradient: in this sense $\chi$ is a bulk index of the horizontal buoyancy flux of horizontal convection. 
The relation~\eqref{justify} also shows that the horizontal flux $J$ is, on average, down the applied surface gradient $\dd \bs/\dd x$. Below in~\eqref{effDiff} we use~\eqref{justify} to define an intuitive ``effective diffusivity'' of HC.

%With~\eqref{justify} we see that the Nusselt number in~\eqref{Sig1} is equivalent to
%\beq
% \Nu = \overline{J \frac{\dd \bs}{\dd x} }\Bigg\slash\, \overline{J_\mathrm{diff} \frac{\dd \bs}{\dd x} }\com
% \label{defNus}
%\eeq
%where $J_\mathrm{diff}(x)$ is the horizontal flux of the diffusive solution $\bdiff$.

\subsection{The Nusselt number of a discontinuous surface buoyancy profile}

There is a fourth  definition of the  HC Nusselt number which, however, only applies to piecewise constant surface forcing profiles, such as 
\beq
\bs(x) = \begin{cases} +\bstar \com \qquad &\text{for $-\ellx/2<x<0$} \, ; \\
-\bstar\com \qquad &\text{for $0<x<+\ellx/2$} \com
\end{cases}
\label{pop}
\eeq
\citep{Shis16,P17}
The advantage of the profile~\eqref{pop} is that the horizontal buoyancy flux through the discontinuity in $\bs(x)$ --- that is $J(0)$ --- is distinguished and provides a ``natural'' definition of the horizontal-convective Nusselt number. 

Now with the discontinuous $\bs(x)$ in~\eqref{pop}, the surface buoyancy gradient is $2 \bstar \delta(x)$ and~\eqref{justify} is particularly simple:
\beq
\chi = - 2 \bstar J(0)/\ellx \per
 \label{simp}
\eeq
For the special $\bs(x)$ in~\eqref{pop} there is a direct connection between $\chi$ and the flux through the location of the discontinuous jump in buoyancy, i.e.~the $\chi$-based definition of $\Nu$ in~\eqref{Sig1} recovers the natural definition of Nusselt number associated with the discontinuous $\bs$ in~\eqref{pop}. Of course, $\Nu$ in~\eqref{Sig1} is not restricted to piecewise constant surface forcing profiles, but can also be applied to smoothly varying $\bs(x)$ such as the sinusoid in~\eqref{sbuoy8}, or to the case with constant $\dd \bs/\dd x$ \citep{R65,SK2011}.

\subsection{Two-dimensional surface buoyancy distributions and the effective diffusivity}

$\Nu$ in~\eqref{Sig1} copes with two-dimensional surface buoyancy distributions, $\bs(x,y)$, such as the examples considered by \cite{Rose17}. In this case, and in analogy with~\eqref{Jdef}, one can define a two-dimensional buoyancy flux, $\bJ$, as the $(z,t)$ average of $(ub - \kappa b_x) \bxh + (v b - \kappa b_y)\byh$. Then it is easy to show that the generalizations of~\eqref{J1} and~\eqref{justify} are
\beq
 h \div \bJ = F\com
 \label{J111}
\eeq
and
\beq
 \overline{\bJ \bcdot \grad \bs } = -\chi \per
 \label{justify2D}
\eeq
Thus the horizontal buoyancy flux $\bJ$ is, on average, down the applied surface buoyancy gradient $\grad \bs$, and $\chi$ emerges as a measure of this horizontal-convective transport.

The down-gradient direction of $\bJ$ suggests a relation between $\bJ$ and
$\grad \bs$ in terms of an ``effective diffusivity'', $D$. Thus we propose that
\beq\label{diffFlux}
 \bJ \approx - D\grad \bs\com
\eeq
where $D$ is a constant. An estimate of $D$ is obtained by minimizing the squared error
\beq
 E(D) \defn \overline{|\bJ+ D \grad \bs|^2}\per
\eeq
 Setting $\dd E/\dd D$ to zero we obtain
\beq
 D \defn - \overline{\bJ\bcdot \grad \bs}\Big \slash \overline{|\grad \bs|^2}\per
 \label{effDiff}
\eeq
Substituting~\eqref{justify2D} into~\eqref{effDiff}, the effective diffusivity, defined via minimization of $E(D)$, is diagnosed as
\begin{align}
 D &= \kappa \laa |\grad b|^2\raa\Big\slash \overline{|\grad \bs|^2}\com  \label{enhanceRat}  \\
  &= \Nu \, \kappa  \la |\grad b_{\text{diff}}|^2\ra \Big\slash \overline{|\grad \bs|^2} \per 
   \label{enhanceRat7}
\end{align}
In~\eqref{enhanceRat} the ratio $\laa |\grad b|^2\raa \slash \, \overline{|\grad \bs|^2}$ emerges as an enhancement factor multiplying the molecular diffusivity $\kappa$ to produce the effective diffusivity $D$. Figure~\ref{Fig3}(b) compares the effective-diffusive flux in~\eqref{diffFlux} against $J(x)$ for the solution shown in figures~\ref{Fig1} and~\ref{Fig2}.

We caution that the good agreement in figure~\ref{Fig3}(b) is only for the sinusoidal $\bs(x)$ in~\eqref{sbuoy8}. We have not tested~\eqref{enhanceRat} using other surface  profiles.
\cite{TsaiSheard2020} introduce a parametrized family of surface buoyancy profiles, with the discontinuous profile~\eqref{pop} as one end member and a profile with uniform buoyancy gradient \citep{R65} as the other.  It would be interesting to systematically  test~\eqref{enhanceRat} using this family.

%Rather than using $J(0)$, \cite{TsaiSheard2020} choose to define the Nusselt number using the $x$ average of $|F|$; at fixed $\Ra$, and with $\Pr=6.14$, they find quantitative variation of $\overline{|F|}$ within this family of profiles. In other words, the numerical value of the Nusselt number, no matter how it is defined, depends on the prescribed $\bs(x)$.

\subsection{Entropy production and surface entropy flux}

Thermodynamics provides a physical interpretation of the definition~\eqref{Sig1} and of the power integral~\eqref{chi7}. In the general equations of fluid mechanics, the rate of entropy generation per unit mass resulting from diffusion of heat is $\beta |\grad T|^2/T^2$, where $T$ is the absolute temperature and $\beta$ is thermal conductivity --- see section 49 of \cite{LLFM}. Within the Boussinesq approximation $T=T_0+T_1$, where $T_0$ is a constant reference temperature, $b = g \alpha T_1$ and $T_0 \gg T_1$; $\beta$ is close to a constant and the diffusivity in the buoyancy equation~\eqref{buoy} is $\kappa = \beta/\rho_0 c_0$, where $c_0$ is a constant heat capacity and $\rho_0$ the constant reference density. Approximating the denominator in $\beta |\grad T|^2/T^2$ with $T_0^2$, the rate of production of entropy per unit volume in a Boussinesq fluid is proportional to $\kappa |\grad b|^2$. Therefore on the left of~\eqref{chi7}, $\chi$ is the volume averaged production of entropy by buoyancy diffusion within the enclosure. The right of~\eqref{chi7} is the non-zero flux of entropy through the top surface which, in statistical steady state, is required to balance the interior entropy production $\chi$.

This thermodynamic balance also applies to  RBC and so the Rayleigh--B\'enard Nusselt number can also be expressed in the form~\eqref{Sig1} \citep{H63, DCIII1996}:  the  Nusselt number $\Nu$ in~\eqref{Sig1} has the ancillary advantage of coinciding with that of Rayleigh--B\'enard and focussing attention on entropy production, $\chi$, as the fundamental quantity determining both the strength of transport and the vigor of mixing in all varieties of convection.

The thermodynamic interpretation of $\chi$ explains why $\chi$ participates in so many identities and inequalities. Alternate Nusselt numbers, such as $\NuF$ in~\eqref{NuF}, do not lead to an analytic framework that can be exploited by variational methods: bounds on the $\Nu$--$\Ra$ relation of horizontal convection  employ~\eqref{Sig1} \citep{SKB04,RBLSY}. Nusselt definitions that use $|F|$ as a device to obtain a non-zero measure of the surface buoyancy flux are arbitrary to the extent that the mean of $F^2$, or indeed any  other functional norm of $F$ or $J$, might be employed; these alternatives to entropy production do  not have a link to the  Boussinesq equations of motion.

\section{Equilibration of the Nusselt number \label{equilibration}}

In this section we summarize the results of a numerical study directed at characterizing the transient adjustment of Nusselt number $\Nu$ in~\eqref{Sig1} to its long-time average. Thus in this section $\Nu(t)$ is the ``instantaneous Nusselt number'' in which $\chi(t)$ is defined via a volume average (with no time averaging). We limit attention to $\Pr=1$ and the sinusoidal $\bs(x)$ in~\eqref{sbuoy8} and discuss both no-slip and free-slip boundary conditions. We consider two-dimensional (2D) solutions with aspect ratios
\beq
 \ell_x/h= 4 \com \qquad \ell_y/h=0\com
\eeq
and three-dimensional (3D) solutions with
\beq
 \ell_x/h= 4 \com \qquad \ell_y/h=1\per
\eeq
We focus on $\Ra = 6.4\times 10^{10}$ --- the same $\Ra$ used in figures~\ref{Fig1} through~\ref{Fig3}. We find no important differences in the equilibration of $\Nu$ between these four cases (no slip versus free slip and 2D versus 3D).

These computations were performed with tools developed by the Dedalus project: a spectral framework for solving partial differential equations \citep[][ \url{www.dedalus-project.org}]{Dedalus2019}. We use Fourier bases in the horizontal, periodic directions and a Chebyshev basis in the vertical, and time-march the spectral equations using a fourth-order implicit-explicit Runge--Kutta scheme. For the 2D solutions the resolution is $n_x\times n_z = 512 \times 128$, and for 3D $n_x\times n_y \times n_z = 512 \times 128\times 128$. We tested the sensitivity of our results by halving this resolution and found only small differences.

\subsection{Equilibration of $\Nu(t)$ and other indices of the strength of convection \label{equilSec}}

\begin{figure}
 \centering
 \includegraphics[width=0.79\textwidth]{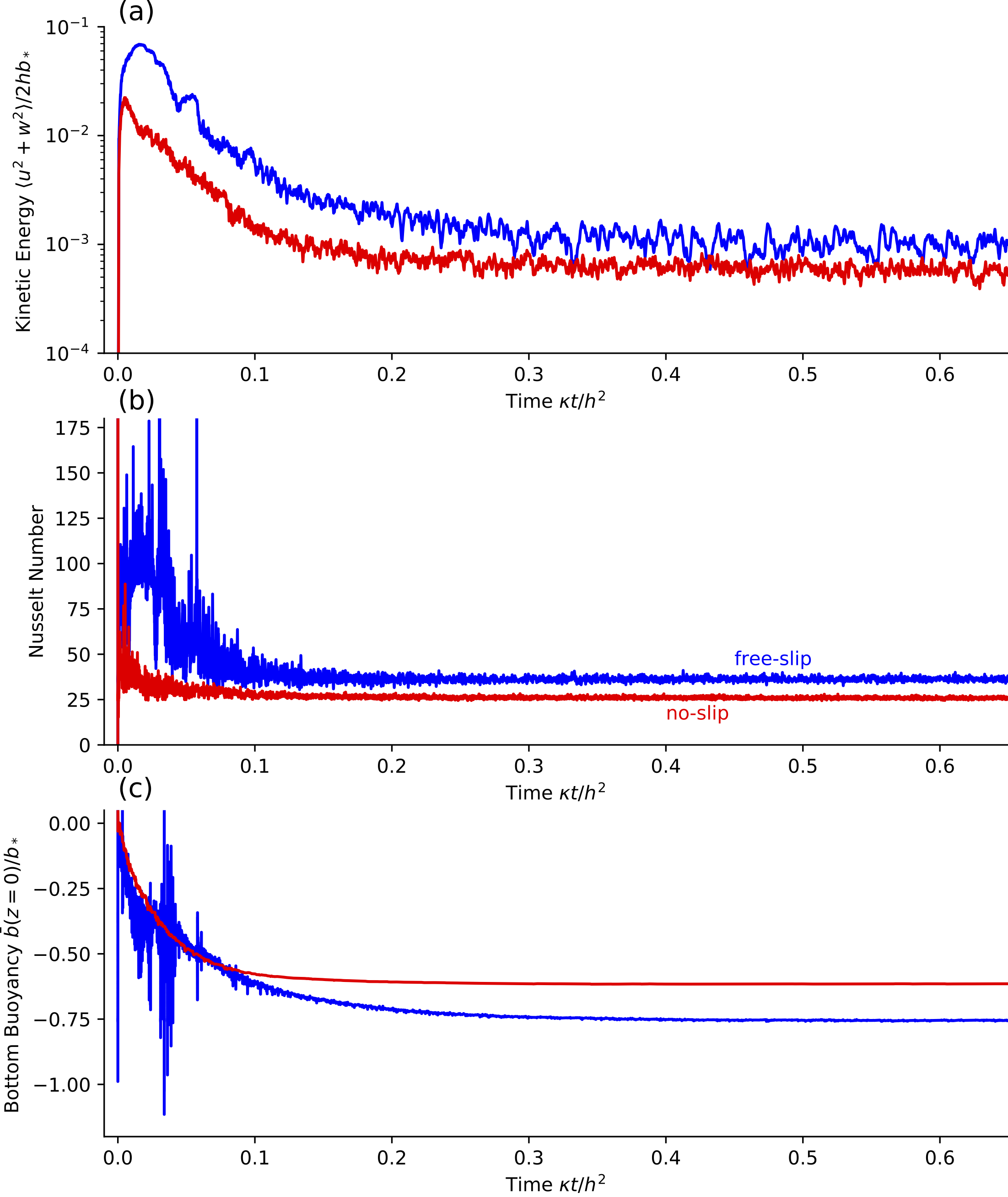}
 \caption{Approach to statistical equilibrium of two cases, one with the no-slip boundary condition (red curves) and the other with free slip (blue curves). Parameters are $\Ra=6.4\times 10^{10}$, $\Pr=1$, $A_x = 4$ and $A_y=0$ (two-dimensional solutions). The initial condition is $\bu=0$ and $b = 0$. (a) Domain-averaged kinetic energy, scaled with $\bstar h$. (b) The ``instantaneous Nusselt number'', $\chi/\chidiff$, with no time-averaging applied to $\chi$. (c) The bottom buoyancy $\bar b(z=0)/\bstar$.}
 \label{Fig4}
\end{figure}

Figure~\ref{Fig4} shows the temporal evolution of the volume averaged kinetic energy, $\Nu$ in~\eqref{Sig1} and the bottom buoyancy $\bar b(z=0)$ of two $\Ra=6.4\times 10^{10}$ solutions: one with no-slip and the other with free-slip boundary conditions. Both solutions in figure~\ref{Fig4} are two-dimensional ($A_y=0$). The
initial condition is $\bu=b=0$, i.e.~the initial buoyancy is equal to the average of $\bs(x)$ in~\eqref{sbuoy8}. The bottom buoyancy in figure~\ref{Fig4}(c) appears as the energy source in~\eqref{PY2.5} and in this sense the free-slip solution, with a larger value of $|\bar b(0)|$, is more strongly forced than the no-slip solution. By all three indices the free-slip flow has a stronger circulation than no-slip.

Both solutions in figure~\ref{Fig4} slowly settle into a statistically steady state with persistent eddying time dependence associated with undulations of the plume that falls from beneath the densest point on the top surface, $z=h$. As noted by \cite{WH2005} and \cite{IV12}, there is an active initial transient during which the flow is much more energetic than its long-time state, which is achieved on the diffusive timescale $h^2/\kappa$. The volume averaged kinetic energy in figure~\ref{Fig4}(a) transiently achieves values more than thirty times larger than the final value at the end of the computation $t = 0.65h^2/\kappa$. But most of this initial excitement subsides by about $t=0.1 h^2/\kappa$ and subsequently there is a slow adjustment lasting until the end of the computation. The kinetic energy and the bottom buoyancy are still slowly decreasing at $t = 0.65h^2/\kappa$. Fortunately, however, the Nusselt number in figure~\ref{Fig4}(b) reaches its final value significantly more rapidly than the other two indices, e.g., beyond about $t=0.15h^2/\kappa$, $\Nu(t)$ is stable. Probably this is because $\Nu(t)$ is determined mainly by transport and mixing in the surface boundary layer, where $|\grad b|$ is largest. The surface boundary layer  comes rapidly into statistical equilibrium. \cite{Griffiths13} and \cite{R98} have previously noted that adjustment of the boundary layer to perturbations in the surface buoyancy is very much faster than the diffusively-controlled adjustment of the deep bulk. From figures~\ref{Fig1} and~\ref{Fig2}, the boundary-layer thickness is about $0.05 h$, hence the diffusive equilibration of the boundary layer occurs on a tiny fraction (1/400) of the vertical diffusive timescale~$h^2/\kappa$.

\subsection{``Cold-start'' initial conditions}

\begin{figure}
 \centering
 \includegraphics[width=.79\textwidth]{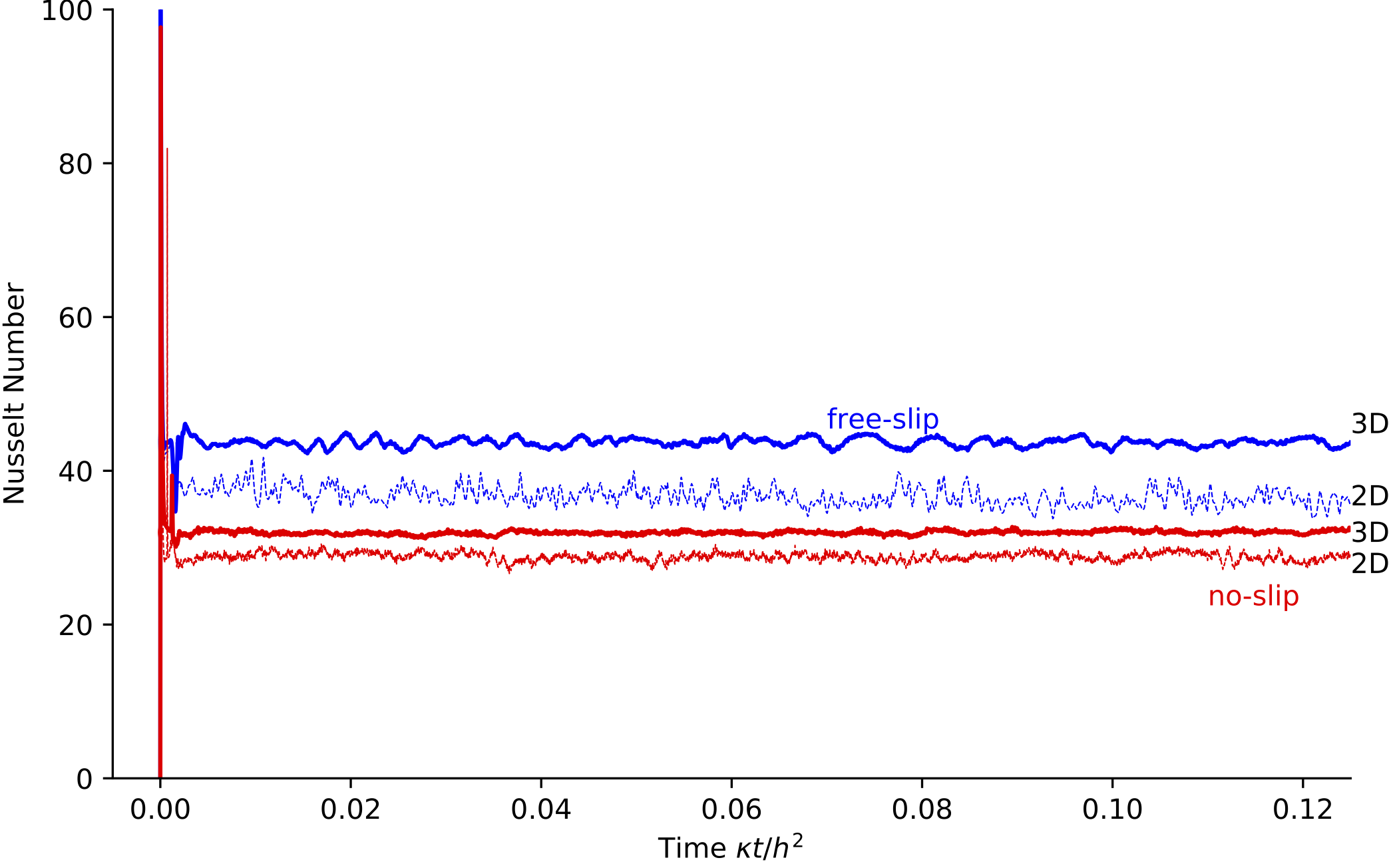}
 \caption{Rapid statistical equilibration of the Nusselt number of four $Ra = 6.4 \times 10^{10}$ solutions with the ``cold-start'' initial condition $b(\bx,t=0) = - 0.74 b_*$.}
 \label{Fig5}
\end{figure}

Numerical resolution of the small spatial scales and fast velocities
characteristic of the initial transient in figure~\ref{Fig4} makes strong demands on both spatial resolution and time-stepping. To reduce the strength of this transient, particularly for 3D integrations with $\Ra$ greater than about $10^9$, we experimented with buoyancy initial condition such as $b(x,y,z,0)=-0.74 \bstar$. This ``cold start'' ensures that the flow begins closer to its ultimate sluggish state, thus rendering the initial transient much less energetic. The cold start makes far less arduous computational demands, both because the weaker transient requires less spatial and temporal resolution and because the $\Nu(t)$ equilibrates even faster than in figure~\ref{Fig4}: see figure~\ref{Fig5}.

\begin{figure}
 \centering
 \includegraphics[width=.79\textwidth]{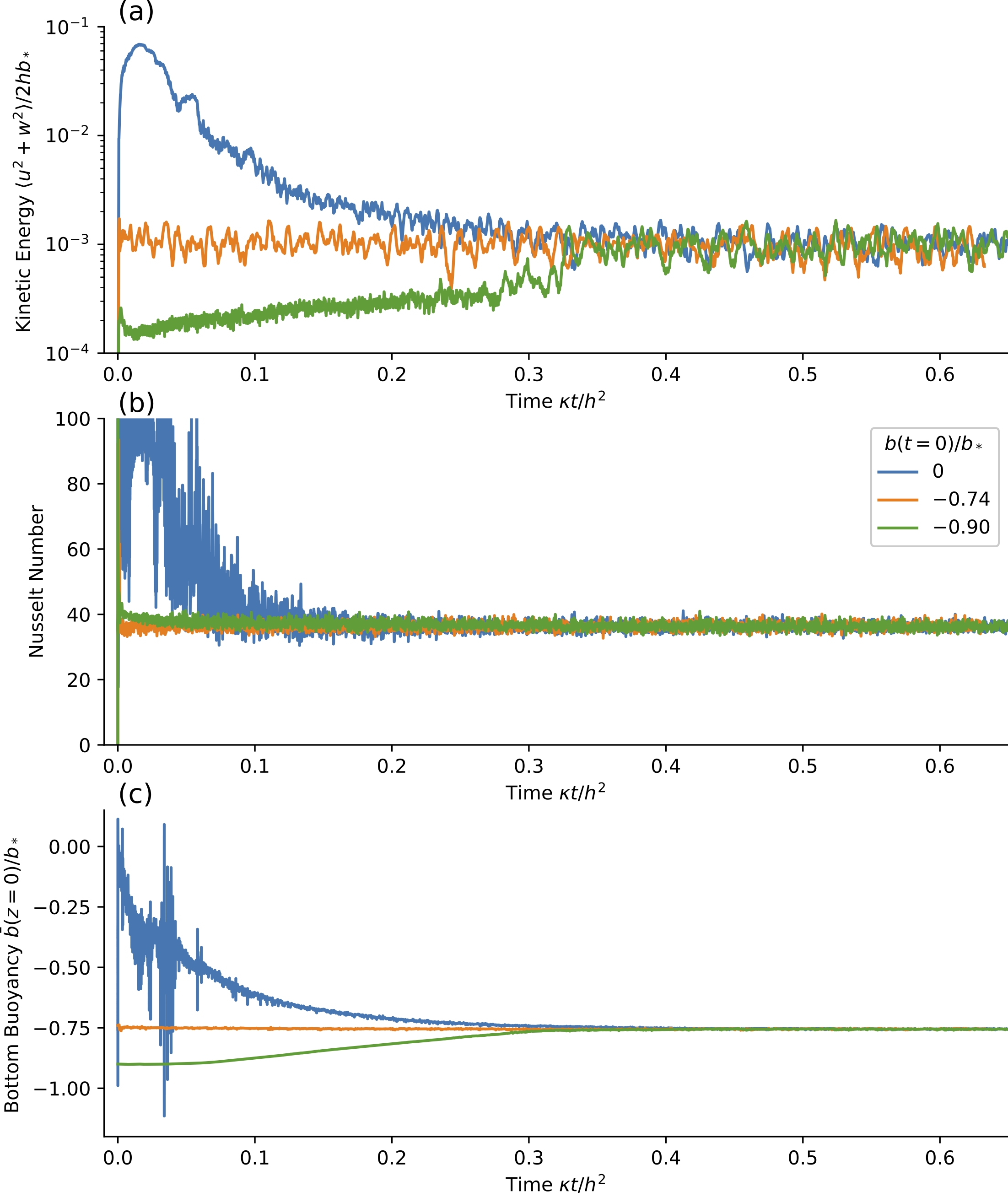}
 \caption{Approach to statistical equilibrium for 2D free-slip simulations at $Ra = 6.4 \times 10^{10}$. We show three initial buoyancy conditions $b(x,z,0) = \bstar\times (0, -0.74,-0.9)$. (a)~Domain-averaged kinetic energy. (b)~The Nusselt number $\Nu(t)$. (c)~The bottom buoyancy $\bar b(z=0)/\bstar$.}
 \label{Fig6}
\end{figure}

In figure~\ref{Fig5} we used the cold initial buoyancy $-0.74 \bstar$ that is
suggested by the long calculation in figure~\ref{Fig4}(c). But usually one must
guess at the initial buoyancy which is closest to the ultimate bottom buoyancy.
The three solutions shown in figure~\ref{Fig6} indicate that the consequences
of a guess that is too cold are not serious. The solution with initial buoyancy
$b(x,z,0)= - 0.9 \bstar$ is too cold: the bottom buoyancy must increase to
about $-0.75 \bstar$ in the long-time state. Nonetheless, the Nusselt number of the too-cold solution in figure~\ref{Fig6}(b) equilibrates quickly. Moreover to estimate $\Nu$ it is better to start too cold than too warm: the too-warm initial condition in figure~\ref{Fig6}, i.e.~$b(x,z,0)= 0$, has a large initial transient in the kinetic energy and $\Nu(t)$ does not stabilize until about $t = 0.2h^2/\kappa$. The kinetic energy and bottom buoyancy in figure~\ref{Fig6}(a) and~(c) require longer evolution than $\Nu(t)$ in order to achieve their final values. Additionally, the transient period for the too-cold initial condition has a much weaker flow compared to the vigorous transient flow of the too-warm solution --- see figure~\ref{Fig6}(a); this reduces the computational effort required to reach statistical steady state.

% \vskip 0.1truein\noindent
%  {\color{red}NCC: OK, to be fair here, our too cold condition is only ~$0.15$ away from the ultimate bottom buoyancy value while our too warm condition is ~$0.75$ away. So it's a bit unfair to compare the two... However, a worthy point is that the ``too cold'' start has a transient phase with low velocities in contrast with the ``too warm'' one; see figure 6(a). This has implications on the time-step required...} \textcolor{Mahogany}{WRY: this might be  true, but I'd like to ignore it. I  won't object if someone else is inspired to modify the paragraph above.} \textcolor{Mahogany}{NCC: I added the red sentence in the end of the paragraph.}
%  \vskip 0.1truein

\begin{figure}
 \centering
 \includegraphics[width=.8\textwidth]{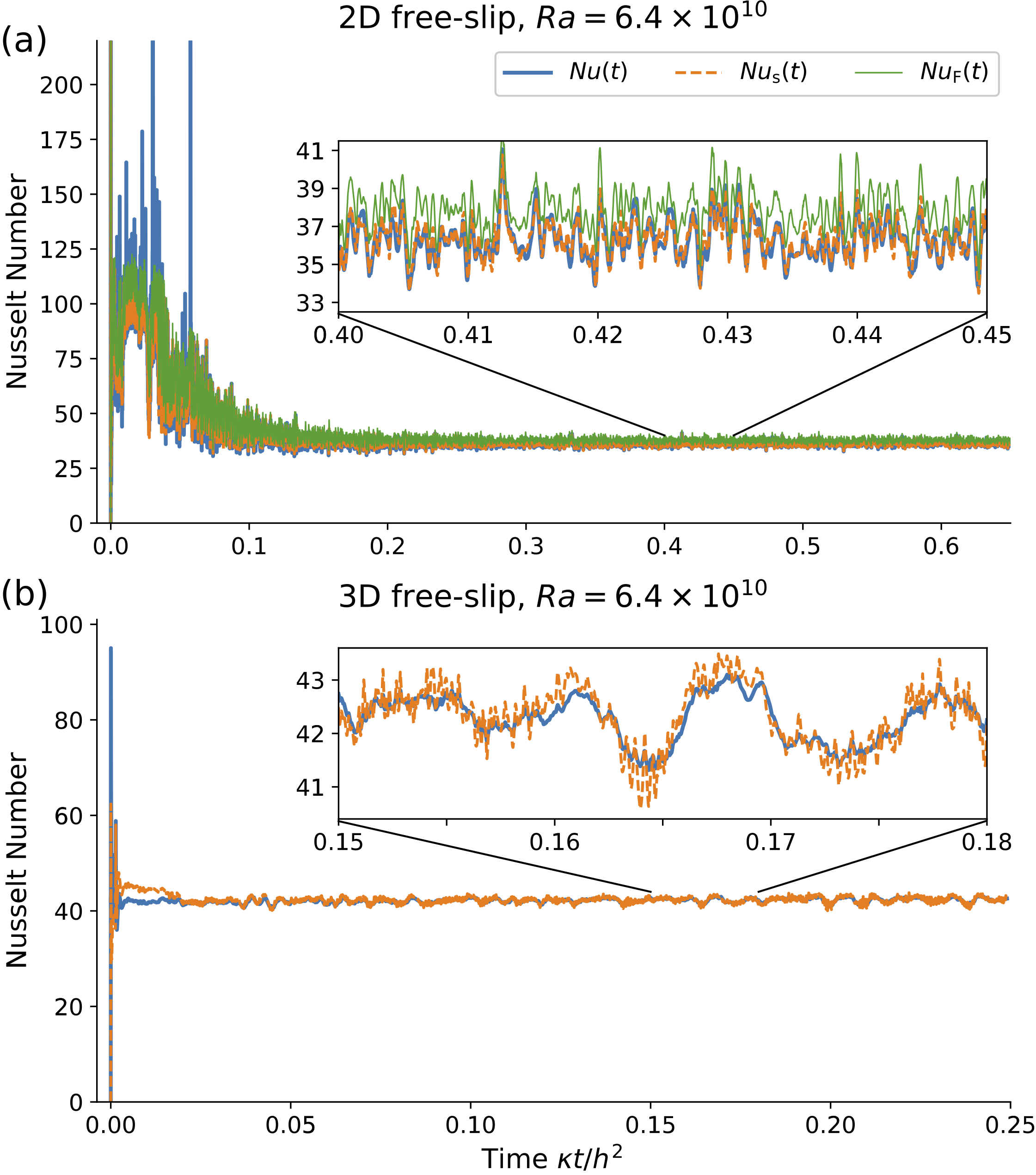}
 \caption{A comparison of the ``instantaneous Nusselt number'' time-series for (a) the 2D free-slip solution at $\Ra = 6.4 \times 10^{10}$ with initial condition $b(x,z,0)=0$ and (b) the 3D free-slip solution at $\Ra = 6.4 \times 10^{10}$ with initial condition $b(x,z,0)=-0.74\bstar$. In both cases $\Pr=1$. $\Nu(t)$ is defined via the volume average of $\kappa |\grad b|^2$, i.e.~the second term on the right of~\eqref{buoyFluc}. The surface Nusselt number, $\Nus(t)$, is defined via the surface average of $\bs$ times the flux $\kappa b_z(x,h,t)$, i.e.~the first term on the right of~\eqref{buoyFluc}. $\NuF(t)$ is defined as  in~\eqref{NuF} without the time average. The time series in panel (b) is from the solution shown in figures~\ref{Fig1} and~\ref{Fig2}.}
 \label{Fig7}
\end{figure}

\subsection{The surface Nusselt number $\Nus$ \label{surfNuss}}

The identity~\eqref{chi7} provides an alternative means of diagnosing the~$\Nu$ by measuring the buoyancy flux $\kappa b_z(x,y,h,t)$ through the top surface of the domain. We refer to this second Nusselt number as the ``surface Nusselt number'', denoted $\Nus(t)$. Multiplying the buoyancy equation~\eqref{buoy} by $b$, and integrating over the volume of the domain, we obtain
 \beq
\frac{\dd}{\dd t} \int \half b^2 \, \dd V = \int_{z=h} \!\!\!\!\! \bs \, \kappa b_z \, \dd A - \kappa \int |\grad b|^2 \, \dd V \per
 \label{buoyFluc}
 \eeq
This shows that the difference between $\Nus(t)$ and $\Nu(t)$ --- the two terms on the right of~\eqref{buoyFluc} --- is related to temporal fluctuations in the domain-integrated buoyancy variance. The buoyancy power integral~\eqref{chi7} is obtained by time-averaging~\eqref{buoyFluc}.

Figure~\ref{Fig7} shows that during the initial transient there are substantial differences between $\Nu(t)$ and $\Nus(t)$. But after the transient subsides, $\Nu(t)$ and $\Nus(t)$ are almost equal, even without time-averaging. This coincidence is most striking in the 2D solution shown figure~\ref{Fig7}(a). For the 3D solution in figure~\ref{Fig7}(b), $\Nus(t)$ closely follows $\Nu(t)$, except for the small high-frequency variability evident in $\Nus(t)$, but not in $\Nu(t)$.

This close agreement between $\Nu(t)$ and $\Nus(t)$ indicates that in  statistical steady state the left hand side of~\eqref{buoyFluc} is a small residual between the two much larger terms on the right. This indicates that the buoyancy boundary layer is strongly controlled by diffusion. Moreover, the source of the low-frequency temporal variability in $\Nu(t)$ and $\Nus(t)$ is the same in the 2D and 3D cases and results from large-scale, dominantly two-dimensional, interior eddies stirring the boundary layer. The high frequency variability evident in the 3D $\Nus(t)$ likely results from the small-scale spanwise ($y$) boundary-layer variability evident in figure~\ref{Fig1}. Although the time series in figure~\ref{Fig7}(b) is from a 3D solution, the spanwise components are weak:
 \beq
 \frac{\la v^2\ra}{\la u^2+ v^2 + w^2\ra} \approx 0.11\com \qquad \text{and} \qquad \frac{\la b_y^2\ra}{\la b_x^2+ b_y^2 + b_z^2\ra} \approx 0.012\per
 \label{weak3D}
 \eeq
Thus, despite the visual impression from figure~\ref{Fig1}, the solution is dominantly 2D with weak spanwise perturbations. To obtain a robustly three-dimensional flow, it seems that~$\Ra$ must be increased above $6.4 \times 10^{10}$ in figure~\ref{Fig1}. This conclusion is  supported by the numerical insensitivity of $\Nu$ to both boundary condition and dimensionality evident in figure~\ref{Fig5}: the time-averaged Nusselt varies by less than a factor of two,  from $25.5$ for no-slip 2D  to $43.9$ for  free-slip 3D. Moreover  the viscous boundary condition has a larger quantitative effect on $\Nu$ than does dimensionality: in figure~\ref{Fig5}  the 2D free-slip solution has larger $\Nu$ than that of the 3D no-slip solution.

In the context of 2D very viscous  ($\Pr = \infty$) HC, \cite{CWHL08}  show that  important  structural features of the flow are independent of boundary condition.  The result in figure~\ref{Fig5} is numerical evidence  that this insensitivity to the viscous boundary condition  extends to 3D HC with $\Pr=1$.

Direct evaluation of $\chi$ via definition~\eqref{chidef} requires the volume integral of $|\grad b|^2$; this integrand  is concentrated on small spatial scales. Thus in  an experimental situation it is likely impossible to evaluate $\chi$ directly from~\eqref{chidef}. The identity~\eqref{buoyFluc} shows that $\chi$ can alternatively be estimated from a surface integral involving the vertical buoyancy flux through the nonuniform surface --- this is the flux of entropy through the surface required to balance interior entropy production associated with molecular diffusion of heat. Figure~\ref{Fig3}(a) indicates that the surface entropy  flux, $\bs F$, is a smooth, large-scale quantity. Thus $\Nus$ is accessible to experimental measurement. Numerical solutions provide both $\Nu$ and $\Nus$ and one can use  this information to test  if the solution is in statistical steady state e.g. as in figure~\ref{Fig7}.

\subsection{The Nusselt number $\NuF$ and its relation to $\Nu$ and $\Nus$ }

In section~\ref{equilSec} we  concluded that the time average of $\Nu(t)$ can be obtained with computations that are significantly shorter than the vertical diffusion time $h^2/\kappa$. This conclusion probably extends to the alternate Nusselt numbers discussed in section~\ref{NussDef}: all these measures of buoyancy flux are designed to diagnose the thickness of the surface  boundary-layer and likely exhibit the rapid equilibration  in  figure~\ref{Fig4}(b). In support of this conclusion, Figure~\ref{Fig7}(a) shows a time series of Rossby's Nusselt number $\NuF(t)$ in~\eqref{NuF}. $\NuF(t)$ is in close agreement with  $\Nu(t)$ and $\Nus(t)$. Moreover, the ratio $\NuF(t)/\Nus(t)$  fluctuates around 1.04 (not shown). 

We were surprised by the close numerical agreement of $\NuF(t)$ with the other Nusselt numbers: \textit{a priori} one anticipates that $\NuF$ should differ from $\Nu(t)$ and $\Nus(t)$ by a nondimensional multiplier. But why is this multiplier close to one? We  can explain this coincidence using the formula for the effective diffusivity in~\eqref{enhanceRat7}. For the sinusoidal buoyancy profile $\bs(x)$ in~\eqref{sbuoy8}, the diffusive solution is
\beq
b_{\text{diff}} = \underbrace{\bstar  \cos (k x)}_{\bs(x)} \,   \frac{\cosh k z}{\cosh kh}\com
\label{diffSol}
\eeq
and so  the effective diffusivity in~\eqref{enhanceRat7}   is
\beq
D = \Nu \, \kappa \,  \frac{\tanh(kh)}{kh}\per
\label{sinuEffDiff}
\eeq 
On the other hand,  from~\eqref{J111} and~\eqref{diffFlux}, the vertical flux is  $F\approx k^2 h D  \bs$; this result can be used to evaluate the numerator in $\NuF$, and the denominator follows from~\eqref{diffSol}:
\begin{align}
\NuF 
&\approx \frac{D}{\kappa} \frac{kh}{\tanh (kh)}\com        \label{effDiff307}            \\
&\approx \Nu \com                                                                  \label{effDiff529}
\end{align}
where~\eqref{sinuEffDiff} was used in passing from~\eqref{effDiff307} to~\eqref{effDiff529}. Note that this result relies on special properties of the sinusoidal $\bs(x)$; for other surface buoyancy profiles $\NuF/\Nu$ will not necessarily be close to one.

%\beq
% \overline{\bJ \bcdot \grad \bs } = -\chi \com
 %\label{justifyAgain}
%\eeq

\section{Conclusion \label{conclusion}}

In section~\ref{NussDef} we discussed four different Nusselt numbers  used as indices of  horizontal-convective heat transport.  We advocate adoption of  $\Nu = \chi/\chidiff$ as the primary index of the strength of HC. This particular Nusselt number is based on $\chi= \kappa\la |\grad b|^2\ra$, which, up to a constant multiplier, is the volume-averaged rate of Boussinesq entropy production. The surface  Nusselt number, $\Nus$,  of section~\ref{surfNuss} is the flux of entropy through the nonuniform surface;  in statistically steady state the surface entropy flux balances interior production. In experimental situations it is easier to measure the surface integral of $\bs F$ required for $\Nus$, than the volume integral of $|\grad b|^2$ demanded by $\Nu$. An advantage of $\chi/\chidiff$ over alternative  HC Nusselt numbers discussed in section~\ref{NussDef} is that the Nusselt number of Rayleigh--B\'enard convection can also be expressed as $\chi/\chidiff$. Thus the thermodynamic interpretation in terms of entropy production provides a unified framework for both varieties of convection. 

The power integral $ \overline{\bJ \bcdot \grad \bs } = -\chi$ provides a connection between  $\Nu$ and the horizontal buoyancy flux $\bJ$. Using this relation  one can introduce the effective diffusivity $D$ in~\eqref{enhanceRat} and~\eqref{enhanceRat7} such that $\bJ \approx - D \grad \bs$. This  establishes a relation between entropy production and horizontal-convective buoyancy  flux. 

In section~\ref{equilibration} we showed that $\Nu(t)$ equilibrates more rapidly than other average properties of HC, such as volume-averaged kinetic energy and bottom buoyancy. With a  cold start the long-term average of $\Nu(t)$ can be estimated with integrations that are much shorter than a diffusive time scale: see figures~\ref{Fig5} and~\ref{Fig6}(b). These numerical results indicate that entropy production, $\chi$, is strongly concentrated in an upper boundary layer and that fast diffusion through this thin layer results in  rapid equilibration of $\Nu$.

%The definition of $\Nu$ in~\eqref{Sig} also applies to Rayleigh--B\'enard convection. Thus the diffusive dissipation of buoyancy variance on the right of~\eqref{justifyAgain} emerges as the prime index of the strength of both varieties of convection.
%

%THANKS
\vspace{1em}
The authors report no conflict of interest. We thank Taimoor Sohail for help with figure~\ref{Fig1} and Basile Gallet and Thomas Bossy for discussion of horizontal convection. We thank the three referees for comments that improved this paper. Computer resources were provided by the Australian National Computational Infrastructure at ANU, which is supported by the Commonwealth of Australia. WRY was supported by the National Science Foundation Award OCE-1657041.

%%%%%%% END %%%%%%%%%
\end{document}